# Some possible explanations of the discrepancies in the results of modelling the leakage current of detectors after hadron irradiation[1]

**S. Lazanu*), I. Lazanu**)**

*) National Institute for Materials Physics, POBox MG-7, Bucharest-Magurele, Romania,
e-mail: lazanu@infim.ro

**) University of Bucharest, Faculty of Physics, POBox MG-11, Bucharest-Magurele, Romania
e-mail: i_lazanu@yahoo.co.uk

## Abstract

In this contribution we argue that the main discrepancies between model calculations and experimental data for leakage current after hadron irradiation could be explained considering the contributions of primary defects in silicon: vacancy, interstitial and $Si_{FFCD}$ defect. The source of discrepancies between data and previous modelling was tentatively attributed to the $Si_{FFCD}$ defect. Vacancies and interstitials have a major contribution to the current short time after irradiation. If these hypotheses are correct, thus, in conditions of continuous long time irradiation, as e.g. LHC and its upgrades in energy and luminosity, S-LHC and V-LHC respectively, these contributions will represent a major problem.

[1] Work in the frame of CERN RD-50 Collaboration

## Introduction

Silicon detectors will be used extensively in experiments at the CERN Large Hadron Collider (LHC) where they will be exposed to fast hadron fluences, as well as in the more hostile environments expected in the machine upgrade in luminosity and energy as SLHC and VLHC [1], [2], [3], [4] respectively. The principal obstacle to long-time operation arises from bulk displacement damage in silicon material, which increases the leakage current in the detector, decreases the satisfactory Signal/Noise ratio, and increases the effective carrier concentration and thus depletion voltage, which ultimately increases the operational voltage of the device beyond the breakdown voltage.

An important old observation consists in the conclusion that there exists a good or reasonable agreement between model and data for the leakage current and effective carrier concentration after lepton or gamma irradiation, and discrepancies up to 2 orders of magnitude (smaller values in model calculation) after hadron irradiation.

In this contribution we argue that the main discrepancies between model calculations and experimental data for leakage current after hadron irradiation could be solved considering the contributions of primary defects in silicon, as vacancies, interstitials and $Si_{FFCD}$, whose existence or characteristics have recently been put in evidence.

## Present status of knowledge about primary defects

The experimental examination of point defects buried in the bulk is difficult and for the various defects is usually indirect. The lattice vacancy and self interstitial are, by their nature, the simplest known defects, produced thermally or by irradiation with energetic particles. In thermal equilibrium the concentration of vacancies and self-interstitials is small because their formation energies are several eV.
The stability of crystalline silicon comes from the fact that each silicon atom can accommodate its four valence electrons in four covalent bonds with its four neighbours. The production of primary defects or the existence of impurities or defects destroys the four-fold coordination.

It has been established that the vacancy takes on five different charge states in the silicon band gap: $V^{2+}$, $V^{+}$, $V^{0}$, $V^{-}$, and $V^{2-}$. Only relatively recently, in a series of theoretical studies [5] and correlated EPR and DLTS experiments of Watkins and co-workers [6] it has been possible to solve some problems associated with the electrical level structure of the vacancy. The charge states $V^{2+}$, $V^{+}$, $V^{0}$ form the so-called negative U system, caused when the energy gain of a Jahn-Teller distortion is larger than the repulsive energy of the electrons, case in which the (0/+) level is inverted in respect to (+/++) level, which are the striking consequence of the fact that the $V^{+}$ charge state is metastable.
For vacancy the structural characteristics are: the bond length in the bulk is 2.35 Å and the bond angle – 109°. The formation energy is 3.01 eV (for p-type silicon), 3.17 eV (intrinsic), 3.14 eV (n-type).

The self-interstitials in silicon could exist in four charge states [7]: $I^{-}$, $I^{0}$, $I^{+}$ and $I^{2+}$.
For interstitials, different structural configurations are possible:

- the hexagonal configuration is a sixfold coordinated defect with bonds of length 2.36 Å, joining it to six neighbours which are fivefold coordinated;
- the tetrahedral interstitial is fourfold coordinated; has bonds of length 2.44 Å joining it to its four neighbours, which are therefore five coordinated;
- the split - <110> configuration: two atoms forming the defect are fourfold coordinated, and two of the surrounding atoms are fivefold coordinated.
- the 'caged' interstitial contains two normal bonds, of length of 2.32 Å, five longer bonds in the range 2.55÷2.82 Å and three unbounded neighbours at 3.10÷3.35 Å.

Recent calculations [8], [9], [10] found that the tetrahedral interstitial and caged interstitial are metastable. For intersitials, the lowest formation energies in eV are 2.80 (for p-type material), 2.98 (for n-type) and 3.31 in the intrinsic case respectively.

New experimental results [7, 11] combined with some old data [12] permitted us to suggest the possible level position assignment for isolated vacancies and interstitials –see Figure 1.

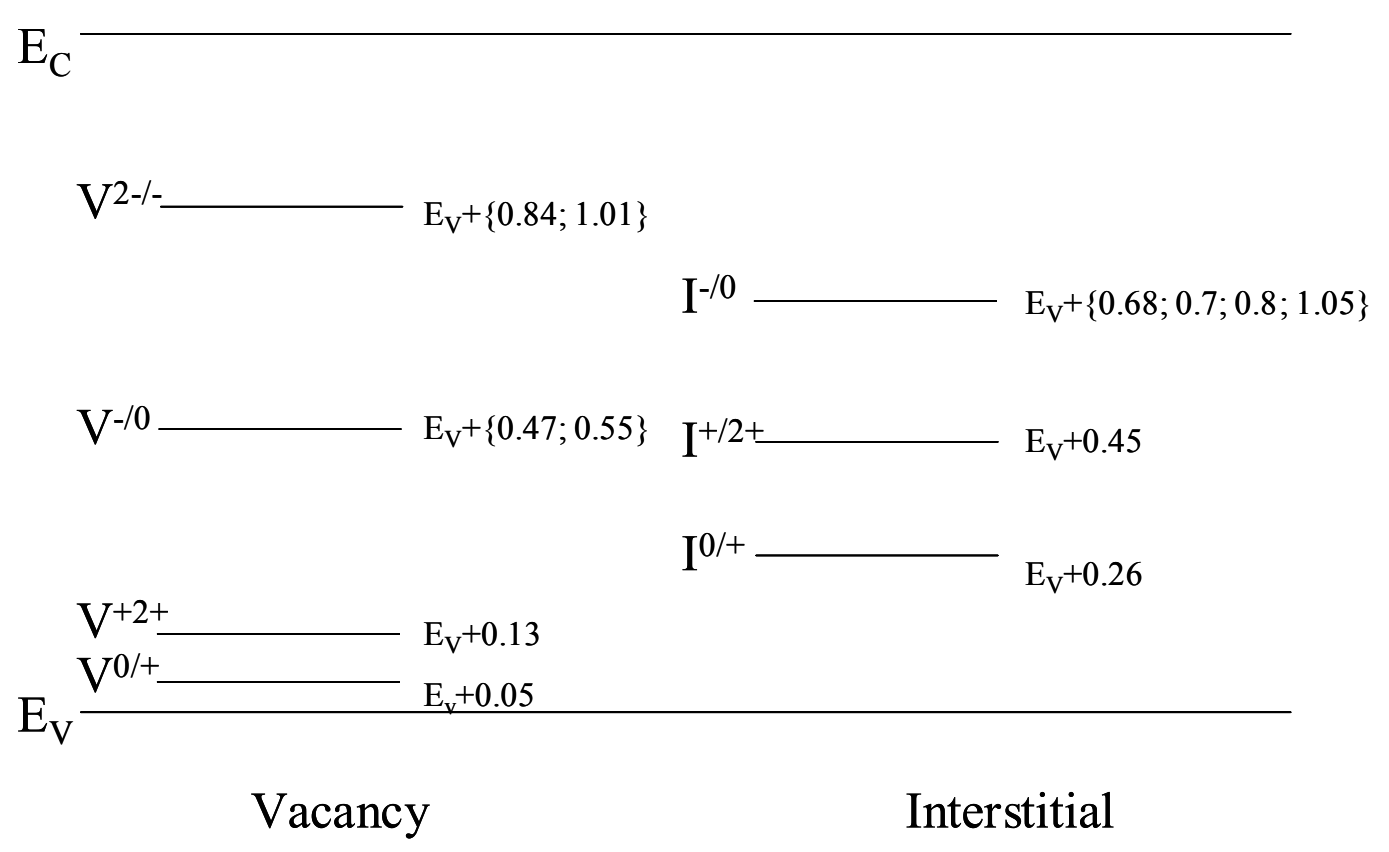

**Figure 1**
Possible level position assignment for isolated vacancies and interstitials

Goedecker and coworkers [13] predicted the existence of a new type of primary defect: $Si_{FFCD}$ (**F**our**f**olded **C**oordinated Silicon **D**efect). It is obtained by moving atoms from the initial positions, but this displacement does not break the bonds with the neighbours. The bound lengths are between 2.25÷2.47 Å and angles vary in the 97÷116° range. The formation energy is 2.45 eV (for p-type silicon), 2.42 eV (intrinsic), 2.39 eV (n-type), lower than the energy of formation of both vacancies and interstitials. The defect has energy levels in the band gap (only calculated) and most probably it is very stable.

## Kinetics of defects – hypothesis of the model and results

Consequence of the irradiation process, primary defects are produces with a rate depending on the incident particle type, flux and energy. This process is supplemented by the thermal rate that is considered in all model calculations.

Thus, the primary reactions considered in the present work are:

$$\begin{aligned} Si \;&\xrightarrow{G}\; V + I \\ &\xrightarrow{G}\; Si_{FFCD} \end{aligned}$$

where: $G = G_{thermal} + G_{irradiation}$

The rate of defect production after irradiation is determined as:

$$G_{irradiation} = \int CPD(E) \times Flux(E) dE$$

where the concentration of primary defects on unit particle fluence (CPD), which is energy and particle dependent, is calculated in the frame of Lindhard's theory [14] and authors' contributions [15]. In Figure 2 a compilation of energy dependencies of the CPD produced by different particles in silicon [16] is presented.

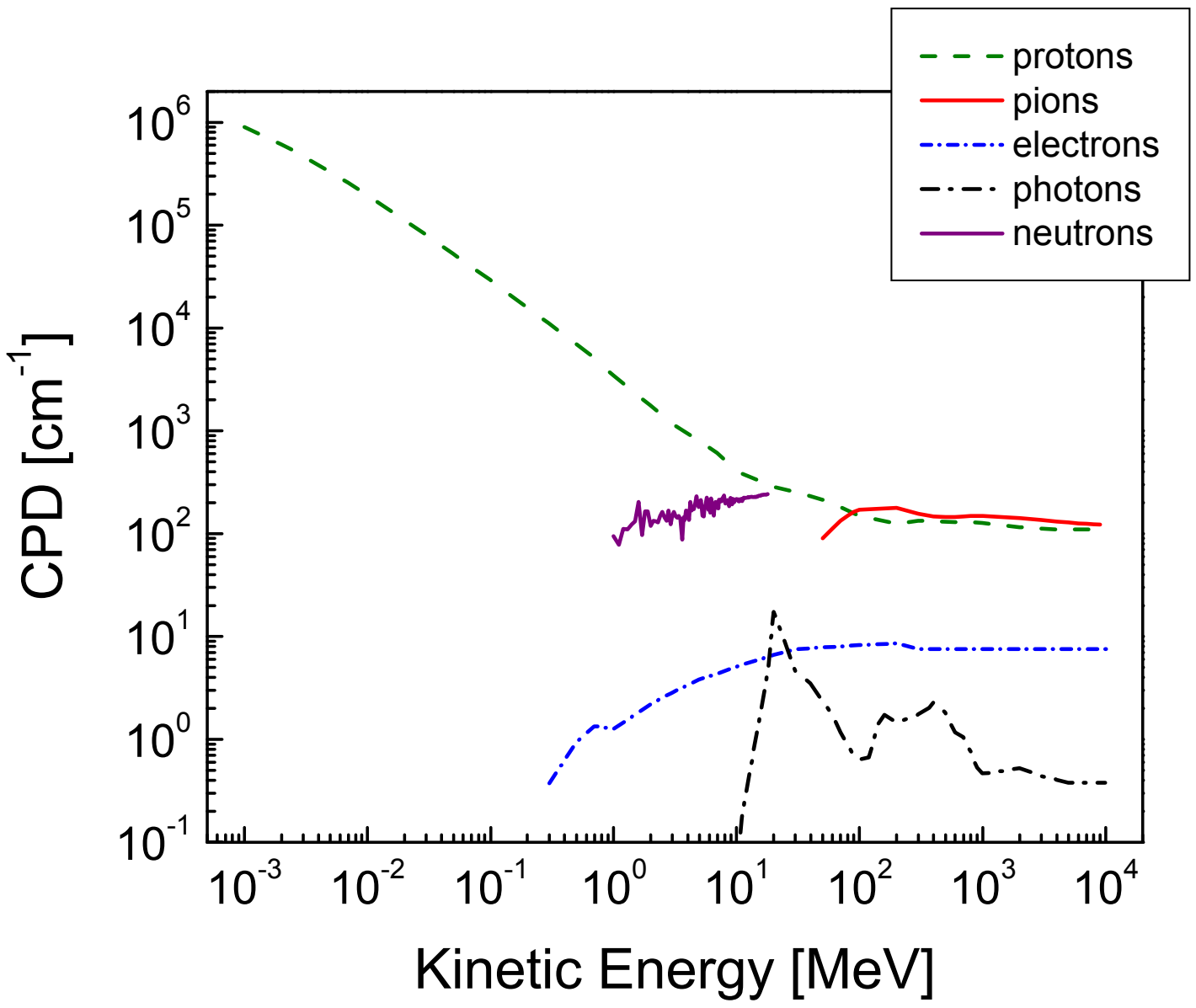

**Figure 2.**
Energy dependence of the concentration of primary defects per unit of fluence produced by different particles in silicon

We supposed that the $Si_{FFCD}$ primary defect is uniformly introduced in the bulk during irradiation and the defect is stable in time. Also we supposed in accord with theoretical predictions that this defect has deep energy level(s) in the gap, probably in the proximity of free vacancy and interstitial.

Vacancies and interstitial recombine, annihilate, migrate to sinks,

$$V + I \rightarrow (VI)_{Frenkel}$$

$$(VI)_{Frenkel} \rightarrow annihilation$$

$$V \rightarrow sinks$$

$$I \rightarrow sinks$$

or interact producing defect complexes:

$$V + V \leftrightarrow V_2$$

$$V + P \leftrightarrow VP$$

$$V + O \leftrightarrow VO$$

$$I + C_s \leftrightarrow C_i$$

$$C_i + C_s \leftrightarrow C_i C_s$$

$$C_i + O_i \leftrightarrow C_i O_i$$

The Frenkel defect is described as a bond defect [17], and not as a two-defect complex. The defect is rather stable at room temperature; where its lifetime is of the order of hours. In the present model calculations, the Frenkel defect, with a deep level supposed in the vicinity of intrinsic level and close to the free vacancy or interstitial levels, has a minor contribution to the leakage current, only short time after irradiation.

In the calculation of the leakage current after irradiation, the SRH model is considered. For defects in different charge states, only the deep levels situated in the vicinity of the intrinsic level have dominant contributions.

In Figures 3a ÷ 3d the time dependence of the alpha constant of the leakage current ($\alpha = \Delta I/(V\Phi)$) is represented as a function of the time after irradiation: points are experimental data and continuous curves – model calculations in the present paper. The effects produced by positive and negative pions, protons and electrons are modelled. A good accord between model calculations and data is obtained in each case.

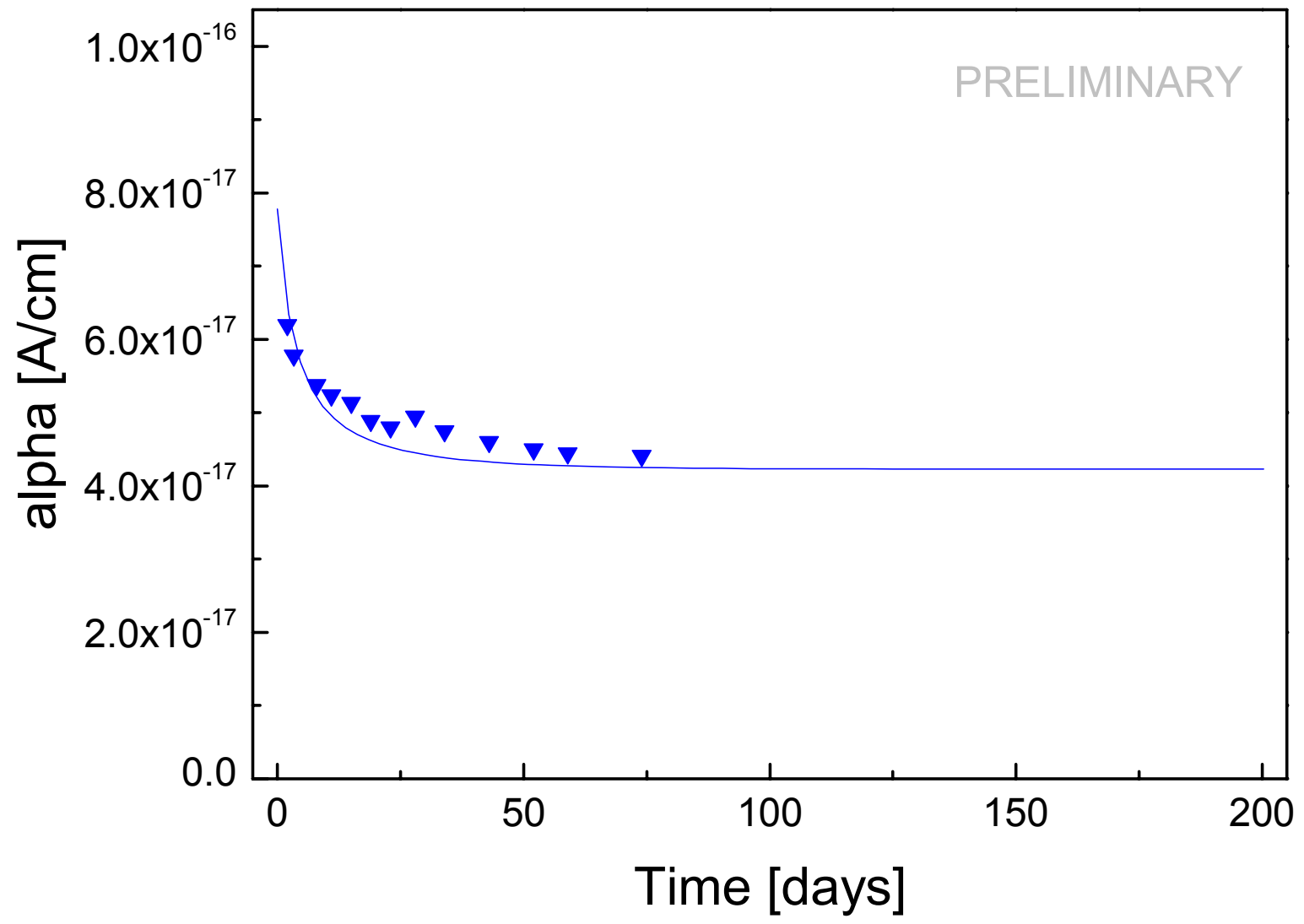

**Figure 3a.**
Time dependence of α degradation constant of the leakage current after positive pion irradiation. Experimental data from Ref. [18]

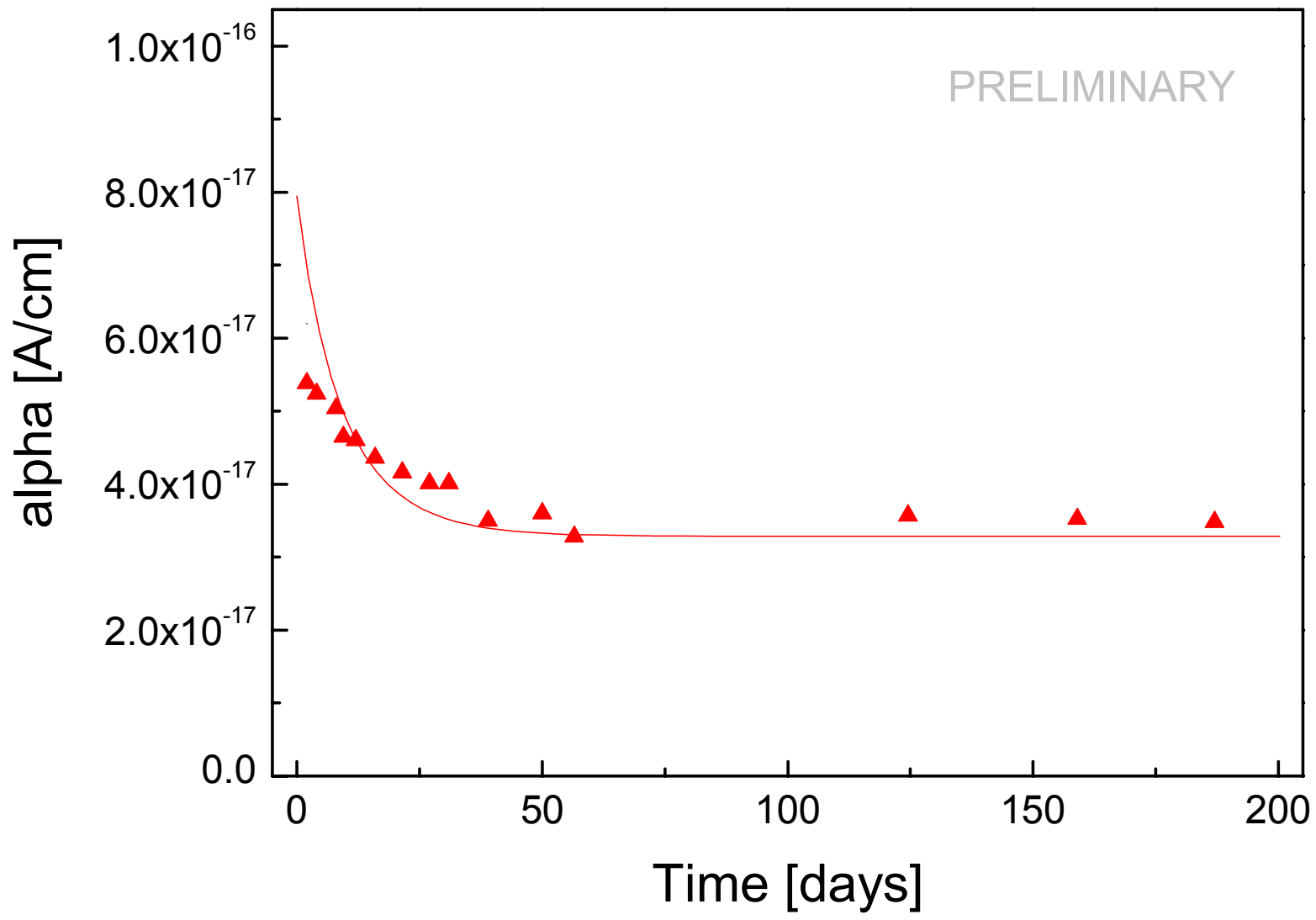

**Figure 3b.**
Time dependence of α after negative pion irradiation. Experimental data from Ref. [18].

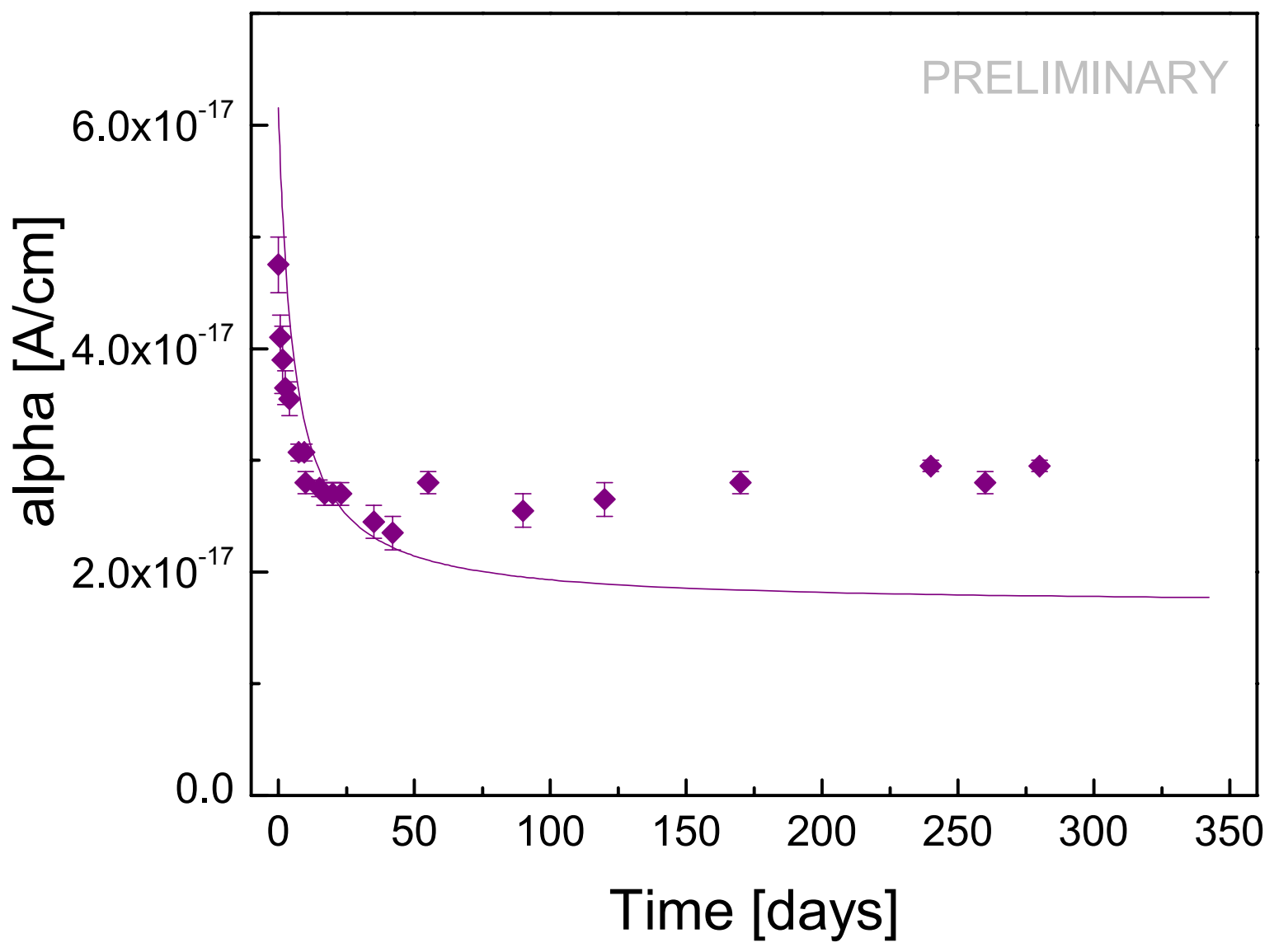

**Figure 3c.**
Time dependence of the α degradation constant after proton irradiation. Experimental data from Ref. 19

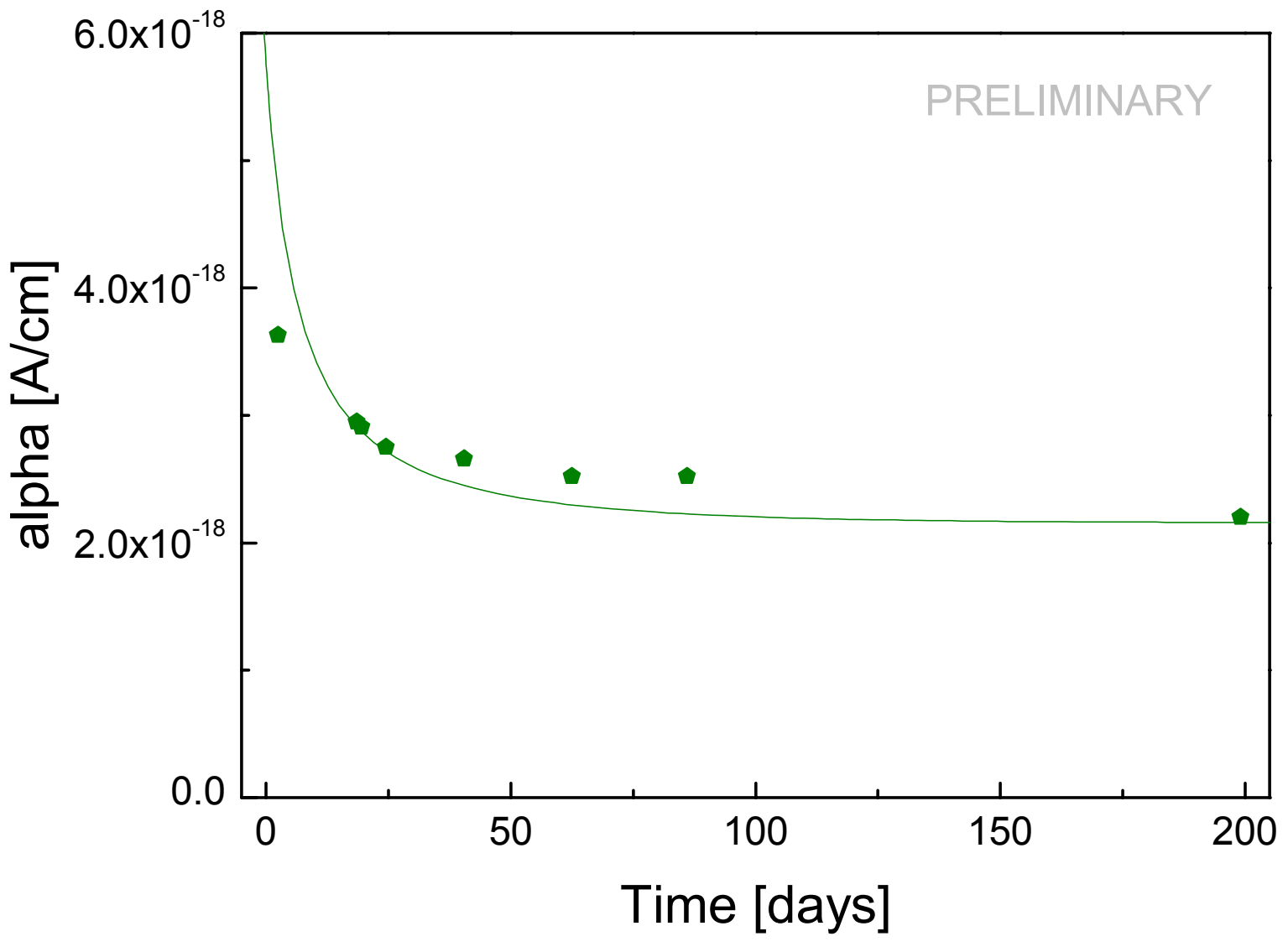

Figure 3d.
Time dependence of the α degradation constant after electron irradiation. Experimental data from Ref. [20]

## Conclusions

In the model hypotheses considered in this work, the preliminary results of the modelling of the leakage current after irradiation with different particles and energies are in very good accord with the data.

The contributions of primary defects to the leakage current were first time considered. In the frame of the model, vacancies and interstitials have a major contribution to the current short time after irradiation. If these hypotheses and results are correct, in conditions of continuous irradiation (as e.g. to LHC and its upgrades) this will represent a major problem.

The existence of a “background” in the leakage current after irradiation, source of discrepancies between data and previous model calculations, was tentatively attributed to the $Si_{FFCD}$ defect. This contribution has been found to be proportional to the concentration of primary defects at the considered fluence If the $Si_{FFCD}$ defect has a deep level in the band gap in the proximity of the intrinsic energy level, as is the case for vacancy and interstitial, thus the concentration of the $Si_{FFCD}$ could be around 10% from all primary defects contribution, but the existence of deep level needs confirmation.

## References

1. U. Bauer et. al., Physics at Futuere hadron Collider, UB-HET-02-01, 2002, http://xxx.lanl.gov/hep-ph/0201227.
2. CERN-ECFA Draft Report of the working group on the future of accelerator-based particle physics in Europe, 2001.
3. F. Gianotti, ICFA, 9/10/2002.
4. G. Ambrosio et. al., Design Study for a Staged Very Large Hadron Collider, Chapter 8: VLHC Experiments and Detectors Issues, Fermilab-TM-2149/2001.
5. G. A. Baraff, E. O. Kane, M. Schlüter, Phys. Rev. Lett. 43 (1979) 956.
6. G. D. Watkins, J. R. Troxell, Phys. Rev. Lett. 44 (1980) 593.
7. В. В. ЛУКЬЯНИЦА, ФТП 37, (2003) 422 or english version: V. V. Lukjanitsa, Semiconductors 37 (2003) 404.
8. R. Needs, J. Phys. Condens. Matter, 11 (1999) 10437.
9. W.-K. Leung, R. J. Needs, G. Rajagopal, S. Itoh, S. Ihara, Phys. Rev. Lett. 83 (1999) 2351.
10. S. J. Clark, G. J. Ackland, Phys. Rev. B 56 (1997) 47.
11. В. В. ЛУКЬЯНИЦА, ФТП 33, (1999) 921.
12. P. M. Fahey, P. B. Griffin, J. D. Plummer, Rev. Mod. Phys. 61 (1989) 289 and references cited therein.
13. S. Goedecker, Th. Deutsch, L. Billard, Phys. Rev. Lett. 88 (2002) 235501.
14. J. Lindhard, V. Nielsen, M. Scharff, P.V. Thomsen, Mat. Phys. Medd. Dan. Vid. Sesk. 33, 1 (1963).
15. S. Lazanu, I. Lazanu, Nucl. Instr. Meth. Phys. Res. A462, 530 (2001).
16. I. Lazanu, S. Lazanu, Nucl. Instr. Meth. Phys. Res. B 201 (2003) 491.
17. F. Cargnoni, C. Gatti, L. Colombo, Phys. Rev. B 57 (1998) 170.
18. J. Bates et al, Nucl. Phys. B. (Proc. Suppl.) 44 (1995) 590.
19. F. Lemeilleur et al., Nucl. Instr. Meth. Phts. Res. A 360 (1995) 438.
20. I. Rashevskaia et al., Nucl. Inst. Meth. Phys. Res. A 485 (2002) 126.